\documentclass[aps, prl, superscriptaddress, preprintnumbers, floatfix, longbibliography,twocolumn,10pt]{revtex4-2}

\usepackage{graphicx,amsfonts,amsmath,amssymb,amstext}
\usepackage{float,wrapfig}
\usepackage{subfigure, psfrag}
\usepackage{dsfont,bm}
\usepackage{color}
\usepackage[colorlinks=true,urlcolor=blue,citecolor=blue,linkcolor=blue]{hyperref}
\hypersetup{breaklinks=true}
\usepackage{verbatim}

\usepackage{array}
\newcolumntype{L}[1]{>{\raggedright\let\newline\\\arraybackslash\hspace{0pt}}m{#1}}
\newcolumntype{C}[1]{>{\centering\let\newline\\\arraybackslash\hspace{0pt}}m{#1}}
\newcolumntype{R}[1]{>{\raggedleft\let\newline\\\arraybackslash\hspace{0pt}}m{#1}}

\definecolor{purple}{rgb}{0.8,0,0.6}
\definecolor{darkgreen}{rgb}{0.00,0.6,0.00}

\usepackage[usenames,dvipsnames]{xcolor}

\extrafloats{100}

\begin{document}

\title{\mbox{Andreev Reflection in Scanning Tunneling Spectroscopy of Unconventional Superconductors}}
\date{May 26, 2023}

\author{P.~O.~Sukhachov}
\email{pavlo.sukhachov@yale.edu}
\affiliation{Department of Physics, Yale University, New Haven, Connecticut 06520, USA}

\author{Felix von Oppen}
\affiliation{\mbox{Dahlem Center for Complex Quantum Systems and Fachbereich Physik, Freie Universit\"{a}t Berlin, 14195 Berlin, Germany}}

\author{L.~I.~Glazman}
\affiliation{Department of Physics, Yale University, New Haven, Connecticut 06520, USA}

\begin{abstract}
We evaluate the differential conductance measured in a scanning tunneling microscopy (STM) setting at arbitrary electron transmission between an STM tip and a two-dimensional (2D) superconductor with arbitrary gap structure. Our analytical scattering theory accounts for Andreev reflections, which become prominent at larger transmissions. We show that this provides complementary information about the superconducting gap structure beyond the tunneling density of states, strongly facilitating the ability to extract the gap symmetry and its relation to the underlying crystalline lattice. We use the developed theory to discuss recent experimental results on superconductivity in twisted bilayer graphene.
\end{abstract}

\maketitle

{\it Introduction.}
The structure of the superconducting order parameter is a defining property of unconventional superconductors~\cite{Mineev-Samokhin:book}. The latter range from high-$T_c$ superconductors such as Ba-doped LaCuO$_3$~\cite{Bednorz-Muller:1986} and BiSrCaCu$_2$O$_x$~\cite{Maeda-Asano:1988} to novel moir\'{e} materials such as twisted bilayer (TBG) and trilayer (TTG) graphene~\cite{LopesdosSantos-CastroNeto:2007,Morell-Barticevic:2010,Bistritzer-MacDonald:2010,Cao-Jarillo-Herrero:2018-TBG,Yankowitz-Dean:2018,Lu-Efetov:2019,Cao-Jarillo-Herrero:2020,Andrei-MacDonald:2020-rev,Oh-Yazdani:2021,Kim-Nadj-Perge:2021} or twisted double-layer copper oxides~\cite{Can-Franz:2020,Volkov-Pixley:2020,Volkov-Pixley:2021,Volkov-Pixley:2022,Zhao-Kim:2021}. In high-$T_c$ superconductors, the large value of the gap allowed one to study its momentum dependence via angle-resolved photoemission spectroscopy (ARPES)~\cite{Damascelli-Shen:review-2003}. The gap symmetry of high-$T_c$ materials was also confirmed by  quasiparticle interference  (QPI)~\cite{Hoffman-Davis:2002,McElroy-Davis:2003}. The much smaller gaps of superconducting TBG and TTG along with the small sample dimensions complicate the use of ARPES, while the large-period moir\'{e} pattern impedes the QPI method. That brings scanning tunneling spectroscopy  (STS) to the fore.

Recent works on TBG and TTG~\cite{Oh-Yazdani:2021,Kim-Nadj-Perge:2021} reveal a V-shaped profile of the differential conductance as a function of bias in the traditional STS regime of weak tunneling (tip relatively far from the sample). This was interpreted as evidence for nodal ($d$-wave) superconductivity. The observation of an enhanced low-bias conductance in the strong-tunneling regime (tip forming a point contact with TBG) was viewed~\cite{Oh-Yazdani:2021} as evidence of Andreev reflection further confirming the unconventional nature of superconductivity in hole-doped TBG.

This experiment prompted us to develop a theory of point-contact tunneling into superconductors with arbitrary gap structures and for arbitrary transmission coefficients of the contact~\footnote{Andreev reflection across a spatially extended (rather than local) junction of a $d$-wave superconductor with a normal metal was considered in Refs.~\cite{Bruder:1990,Tanaka-Kashiwaya:1996,Kashiwaya-Tanaka:2000}.}. As tip-sample tunneling does not conserve momentum, it is difficult to reconstruct the gap structure solely from the differential conductance in the weak-tunneling regime. In this regime, the differential conductance yields the energy dependence of the tunneling density of states, which carries some information on the momentum dependence of the absolute value of the gap. Our theory provides access to considerably more extensive information, including the gap symmetry, by synthesizing data taken in the weak- and strong-tunneling regimes. The additional information enters through the phase sensitivity of Andreev reflections, which dominate STS data in the strong-tunneling limit~\cite{Ruby-Franke:2015}.

{\it Scattering matrix formalism for STM tip.}
We view the contact between tip and 2D system as a single-mode quantum point contact opening into a (super)conducting sheet of material. For a pointlike tip and assuming time-reversal symmetry (TRS) of the normal state so that $s(\varepsilon)=s^{T}(\varepsilon)$, the contact can be described by the two-channel scattering matrix
\begin{equation}
\label{S-def}
s(\varepsilon) =\left(
  \begin{array}{cc}
    s_{0}^{\prime}(\varepsilon) & t(\varepsilon) \\
    t(\varepsilon) & s_{0}(\varepsilon) \\
  \end{array}
\right).
\end{equation}
The amplitude $s_0^\prime(\varepsilon)$ describes reflection between incoming and outgoing channels in the tip, and the transmission amplitude $t(\varepsilon)$ controls the differential conductance of the contact in the normal state, $G_n(V)=G_Q|t(eV)|^2$~\footnote{Microscopically, $|t(\varepsilon)|^2$ depends on the distance between the tip and the 2D material and the local density of states at the point of tunneling ${\bf r}_0$.}. Here $G_Q=e^2/(\pi \hbar)$ is the conductance quantum.

A pointlike tip couples to a single channel of the 2D system, so that  scattering between in- and outgoing waves in the 2D system is described by the S-matrix element $s_0(\varepsilon)$. For a uniform system, $s_0(\varepsilon)$ describes scattering in the zero angular momentum channel: an arbitrary incoming wave $\psi^{\rm in}$ in the substrate is scattered into the outgoing wave $\psi^{\rm out}=[(\hat{I}-\hat{P})+s_0(\varepsilon)\hat{P}]\psi^{\rm in}$, with $\hat P$  projecting onto zero angular momentum. A reflectionless junction between tip and system has $s_0(\varepsilon)=0$, while $|s_0(\varepsilon)|=1$ in the absence of tunneling.

The generalization to 2D crystals modifies the projection operator $\hat P$. For a given dispersion relation $\xi({\bf k})$ (measuring energies from the Fermi energy), the wave vectors ${\bf k}$ at a given energy $\varepsilon$ are defined by $\xi({\bf k})=\varepsilon$. The angular distribution is governed by the Bloch function $u_{\bf k}({\bf r}_0)$ at the position ${\bf r}_0$ of STM tip, so that the projection onto the single channel of the system is effected by the operator
\begin{equation}
\label{Phat}
\hat{P}\psi^{\rm in}_{\mathbf{k}} = u_{\mathbf{k}}({\bf r}_0)\!\!\!
\sum_{\xi({\mathbf{k}}^\prime)=\varepsilon}\!\!\! u_{\mathbf{k}^\prime}^*({\bf r}_0)\psi^{\rm in}_{\mathbf{k}^\prime}\equiv u_{\mathbf{k}}({\bf r}_0)\langle u_{\mathbf{k}^\prime}^*({\bf r}_0)\psi^{\rm in}_{\mathbf{k}^\prime}\rangle_\varepsilon
\end{equation}
with a properly normalized $u_{\mathbf{k}}({\bf r}_0)$. Here, $\langle\dots\rangle_\varepsilon$ stands for averaging over the constant-energy contour.

For contacts between normal-metal tip and superconductor, we extend the scattering matrix to Nambu space, using $s^{*}(-\varepsilon)$ instead of $s(\varepsilon)$ for holes~\cite{Beenakker:1992,Nazarov-Blanter:book}. Below, we exploit the particle-hole symmetry to focus on positive energies $\varepsilon>0$. We also neglect the energy dependence of $s(\varepsilon)$, assuming it to be featureless for energies of the order of the gap $|\Delta|$.

\begin{figure}[t]
\centering
\subfigure{\includegraphics[width=0.99\columnwidth]{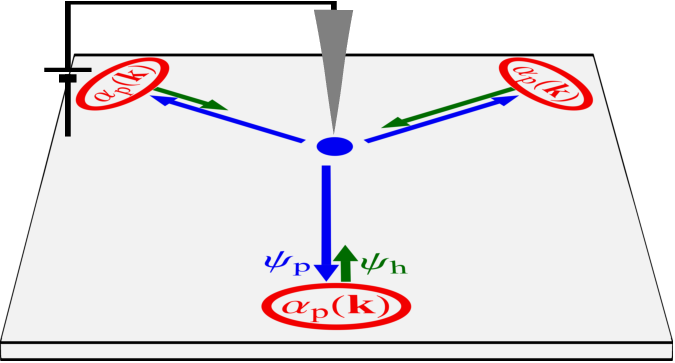}}
\caption{Electron transport in a setup where an STM tip is placed over a high-symmetry point of a 2D superconductor. Symmetric blue arrows: the particle wave spreading from the tip carries the symmetry of crystalline lattice. Asymmetric green arrows: the Andreev-reflected hole wave [Eq.~(\ref{psih1})] carries information about the superconducting gap symmetry, which may differ from the crystalline one.
}
\label{fig:STM-Andreev}
\end{figure}

{\it Andreev and normal reflection.} An electron tunneling into the 2D sample forms an expanding particle wave $\psi_{p\mathbf{k}}^{(1)}$ with amplitude $t$ and directional profile determined by the Bloch function, $\psi_{p\mathbf{k}}^{(1)}= t\,u_{\mathbf{k}}({\bf r}_0)$. The superconductor retroreflects the particle into a counterpropagating hole~\cite{Andreev:1964}; see Fig.~\ref{fig:STM-Andreev} for a sketch. When the coherence length is larger than  the Fermi wavelength, we can account for Andreev reflection within the eikonal approximation: the Andreev amplitude $\alpha(\mathbf{k},\varepsilon)$ depends on the superconducting gap $\Delta(\mathbf{k})$ at the same wave vector $\mathbf{k}$ allowing us to use the result of Refs.~\cite{Beenakker:1992,Nazarov-Blanter:book} at each $\mathbf{k}$,
\begin{equation}
\alpha_{p,h}(z)=\exp\left(\pm i\arg{z}-i\arccos |z|\right), \, z(\mathbf{k},\varepsilon)=\frac{\varepsilon}{\Delta(\mathbf{k})}.
\label{alpha}
\end{equation}
Here, $+(-)$ corresponds to $p\to h$ $(h\to p)$ conversion~\footnote{We will suppress the argument $\varepsilon$ of $\alpha_{p,h}$ in the intermediate equations.}. We restrict considerations to a spin-singlet or polarized spin-triplet superconductor, so $\Delta(\mathbf{k})$ is viewed as a scalar. The analytical continuation to $|z|>1$ is determined by the requirement $|\alpha_{p,h}|\leq 1$.

The gap anisotropy becomes imprinted in the retroreflected wave, $\psi^{\phantom{1}}_{h\mathbf{k}}=\alpha_p(\mathbf{k})\psi_{p\mathbf{k}}^{(1)}$. Only part of it, $\hat{P}\psi^{\phantom{1}}_{h\mathbf{k}}$,  scatters off the tip, while the complement, $(\hat{I}-\hat{P})\psi^{\phantom{1}}_{h\mathbf{k}}$, is oblivious to its presence. Thus, the hole escapes into the tip with amplitude $t^{*}\langle u_{\mathbf{k}}^{*}({\bf r}_0)\alpha_p(\mathbf{k})\psi_{p\mathbf{k}}^{(1)}\rangle_\varepsilon$. The part of the hole wave $\psi^{\phantom{1}}_{h\mathbf{k}}$, which remains within the 2D material,  takes the form $\psi_{h\mathbf{k}}^{(1)}= [(\hat{I}-\hat{P})+s_0^*\hat{P}]\psi_{h\mathbf{k}}$, {\sl i.e.},
\begin{equation}
\psi_{h\mathbf{k}}^{(1)} = \left[\hat{I}-\left(1-s_0^*\right)\hat{P}\right]\alpha_p(\mathbf{k})\psi_{p\mathbf{k}}^{(1)},\,\, \psi_{p\mathbf{k}}^{(1)}= t\,u_{\mathbf{k}}({\bf r}_0)\,.
\label{psih1}
\end{equation}
Retroreflection of the hole wave reconverts it into a particle wave, $\alpha_h(\mathbf{k})\psi_{h\mathbf{k}}^{(1)}$. Similar to the hole, the particle splits between the tip and the 2D material with amplitudes $t\langle u_{\mathbf{k}}^{*}({\bf r}_0)\alpha_h(\mathbf{k})\psi_{h\mathbf{k}}^{(1)}\rangle_\varepsilon$, and
\begin{equation}
\psi_{p\mathbf{k}}^{(2)} = \left[\hat{I}-\left(1-s_0\right)\hat{P}\right] \alpha_h(\mathbf{k})\psi_{h\mathbf{k}}^{(1)}\, ,
\label{psie2}
\end{equation}
respectively. Then the process repeats: $\psi_{p\mathbf{k}}^{(2)}$ is retroreflected into a hole; the hole is partially absorbed into the tip with amplitude $t^*\langle u_{\mathbf{k}}^{*}({\bf r}_0)\alpha_p(\mathbf{k})\psi_{p\mathbf{k}}^{(2)}\rangle_\varepsilon$ and partially scattered off it. Summing over cycles, we obtain the full Andreev-reflection ($r_{ph}$ and $r_{hp}$) and normal-reflection ($r_p$ and $r_h$) amplitudes. For example,
\begin{eqnarray}
r_{ph}&=&|t|^2\left\langle u_{\mathbf{k}}^{*}({\bf r}_0)\sum_{n=0}^\infty \hat{L}^n\, \alpha_{p}(\mathbf{k})u_{\mathbf{k}}({\bf r}_0)\right\rangle^{\phantom{\dagger}}_\varepsilon\,,\label{rA}\\
\hat{L}&\equiv&\alpha_p(\mathbf{k})\left[\hat{I}-(1-s_0)\hat{P}\right]\alpha_h(\mathbf{k})\left[\hat{I}-(1-s_0^*)\hat{P}\right]\,.
    \nonumber
\end{eqnarray}
Symbolically performing the summation in Eq.~(\ref{rA}) gives $r_{ph}=|t|^2\langle u_{\mathbf{k}}^{*}({\bf r}_0){M}(\mathbf{k})\rangle^{\phantom\dagger}_\varepsilon$ with
\begin{equation}
M(\mathbf{k})=(\hat{I}-\hat{L})^{-1}\alpha_p(\mathbf{k})u_{\mathbf{k}}({\bf r}_0).
\label{Mphi}
\end{equation}
We recast Eq.~(\ref{Mphi}) as the integral equation
\begin{equation}
(\hat{I}-\hat{L}) M(\mathbf{k})=\alpha_p(\mathbf{k})u_{\mathbf{k}}({\bf r}_0)\,.
\label{intMphi}
\end{equation}
Since the operator $\hat{L}$ has a separable kernel~\cite{Morse-Feshbach:book}, we solve Eq.~(\ref{intMphi}) by standard means~\cite{SM} and express $M(\mathbf{k})$ in terms of three parameters:
\begin{eqnarray}
\label{aph2}
a_{p,h} &=& \left\langle \left|u_{\mathbf{k}}({\bf r}_0)\right|^2 \frac{\alpha_{p,h}(\mathbf{k},\varepsilon)}{1-\alpha_p(\mathbf{k},\varepsilon)\alpha_h(\mathbf{k},\varepsilon)} \right\rangle_\varepsilon,\\
\label{aeh2-def}
a_{ph} &=& \left\langle \left|u_{\mathbf{k}}({\bf r}_0)\right|^2
\frac{\alpha_p(\mathbf{k},\varepsilon)\alpha_h(\mathbf{k},\varepsilon)}{1-\alpha_p(\mathbf{k},\varepsilon)\alpha_h(\mathbf{k},\varepsilon)}\right\rangle_\varepsilon.
\end{eqnarray}
Here, we restored the energy argument in $\alpha_{p,h}(\mathbf{k},\varepsilon)$. The averaging $\langle|u_{\mathbf{k}}({\bf r}_0)|^2\dots\rangle_\varepsilon$ is defined by
\begin{equation}
\left\langle \left|u_{\mathbf{k}}({\bf r}_0)\right|^2\dots\right\rangle_\varepsilon= \frac{\int d^2k\,\delta (\xi({\mathbf{k}})-\varepsilon)\left|u_{\mathbf{k}}({\bf r}_0)\right|^2\dots}{\int d^2k\, \delta (\xi({\mathbf{k}})-\varepsilon)\left|u_{\mathbf{k}}({\bf r}_0)\right|^2}.
\end{equation}
Using the explicit form~\cite{SM} of $M(\mathbf{k})$ in the expression for $r_{ph}$, we find the Andreev-refection amplitude
\begin{equation}
\label{rAfinal}
r_{ph} = \frac{|t|^2a_p}{1 + \left(2-s_{0}-s_{0}^{*}\right)a_{ph} +\left|1-s_{0}\right|^2 \left(a^2_{ph} -a_p a_h\right)}.
\end{equation}
Similarly, the normal-reflection amplitude is
\begin{equation}
\label{rnfinal}
r_p = s_0^{\prime} +\frac{t^2\left[a_{ph} +\left(1-s_{0}^{*}\right) \left(a_{ph}^2 -a_p a_h\right)\right]}
{1 + \left(2-s_{0}-s_{0}^{*}\right) a_{ph} +\left|1-s_{0}\right|^2 \left(a_{ph}^2 -a_p a_h\right)}.
\end{equation}
The amplitudes $r_{hp}$ and $r_{h}$ are obtained from Eqs.~(\ref{rAfinal}) and (\ref{rnfinal}) by replacing $a_p\leftrightarrow a_h$, $s_0\leftrightarrow s_0^{*}$, and $t\leftrightarrow t^{*}$. Because of the unitarity of the scattering matrix Eq.~(\ref{S-def}), $r_{ph}$ and $|r_p|$ depend only on a single matrix element $s_0$; its magnitude (but not phase) is fixed by  $G_n/G_Q\equiv |t|^2=1-|s_0|^2$.

The Andreev- and normal-reflection amplitudes in Eqs.~(\ref{rAfinal}) and (\ref{rnfinal}) depend on the energy $\varepsilon$ of the incoming electron via Eqs.~(\ref{aph2}) and (\ref{aeh2-def}). The information on the gap structure $\Delta(\mathbf{k})$ and the crystal symmetry is encoded, respectively, in the $\mathbf{k}$-dependences of the retroreflection amplitudes Eq.~(\ref{alpha}) and the Bloch functions $u_{\mathbf{k}}({\bf r}_0)$.

{\it Differential conductance.} We can now express the differential conductance
$ G(V) = {dI(V)}/{dV}$ in terms of the amplitudes $r_{ph}$ and $r_p$. For $V>0$, one has~\cite{BTK:1982}
\begin{equation}
\label{dIdV}
G(V,\mathbf{r}_0) = G_Q \left[1 - \left|r_p(eV,\mathbf{r}_0)\right|^2+\left|r_{ph}(eV,\mathbf{r}_0)\right|^2\right].
\end{equation}
The conductance for $V<0$ follows by replacing $r_p(eV,\mathbf{r}_0)\to r_h(-eV,\mathbf{r}_0)$ and $r_{ph}(eV,\mathbf{r}_0)\to r_{hp}(-eV,\mathbf{r}_0)$.

\begin{figure*}[!ht]
\centering
\includegraphics[width=0.99\textwidth]{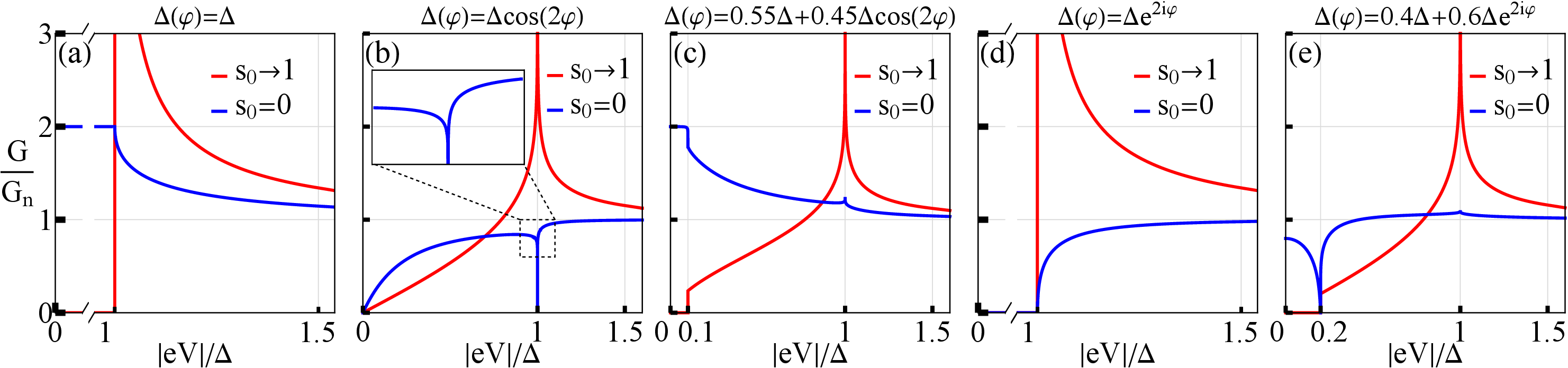}
\caption{
Dependence of the normalized differential conductance $G/G_{n}$ on bias $V$ for weak ($s_0\to 1$, red) and strong ($s_0=0$, blue) tunneling at a high-symmetry point. The conductance is evaluated with the help of Eqs.~(\ref{aph2})--(\ref{dIdV}) for a 2D superconductor with a circular Fermi surface (parameterized by the angle $\varphi$) and gap $\Delta(\mathbf{k})= \Delta(\varphi)$.
(a): $s$-wave superconductor;
(b): $d$-wave superconductor preserving time-reversal symmetry; $G(V)$ remains linear in the limit $V\to 0$ at any tunneling strength; the van Hove singularity at $s_0\to 1$ is replaced by a Fano resonance (inset) at strong tunneling;
(c): s+d gap preserving time-reversal symmetry, but breaking the lattice point symmetry; parameters chosen to preserve the $G(0)=2G_Q$, but significantly shrink the plateau $G(V)<2G_Q$ at $V>0$ compared to the case of $s$-wave superconductor ({\sl cf.} (a));
(d): d+id gap preserving point-group symmetry, but breaking time-reversal symmetry; $G(V)$ remains zero below the gap at any $s_0$; (e): s+d+id gap breaking point-group and time-reversal symmetries; a prominent Fano resonance develops at $eV={\rm min}\{|\Delta(\varphi)|\}$ in the strong tunneling limit.
}
\label{fig:G}
\end{figure*}

Equations~(\ref{aph2})--(\ref{dIdV}) provide a highly flexible framework for describing local tunneling spectroscopy of 2D superconductors and constitute the main advance of this work. They account for arbitrary superconducting gaps as well as the band structure, covering the entire crossover from weak to strong tunneling between tip and superconductor. While the weak-tunneling regime probes the local tunneling density of states, the strong-tunneling regime is dominated by Andreev processes, providing complementary information about the superconducting order parameter. Below, we illustrate the utility of our approach by focusing on several characteristic limits.

In the weak-tunneling limit $s_0\to 1$, the differential conductance is governed by the tunneling density of states $\nu(eV)$ of the superconductor. Indeed, for $s_0\to 1$, only the term $\propto a_{ph}$ in Eq.~(\ref{rnfinal}) contributes, so that Eq.~(\ref{dIdV}) reduces to $G(V)=G_n\nu(eV)/\nu_n$ (with the tunneling density of states $\nu_n$ of the normal state). A fully gapped anisotropic superconductor with $\min\{|\Delta ({\mathbf{k}})|\}=\Delta_{\rm min}$ is signaled by zero conductance in the interval $|eV|<\Delta_{\rm min}$; see, {\sl e.g.}, Figs.~\ref{fig:G}(a), \ref{fig:G}(c)--\ref{fig:G}(e). In contrast, a nodal point in $\Delta(\mathbf{k})$ results in a V-shape profile $G(V)\sim G_n|eV|/\Delta$ at low biases; see Fig.~\ref{fig:G}(b); hereinafter $\Delta$ is the characteristic value of $|\Delta ({\mathbf{k}})|$. Apart from this distinction, weak-tunneling data do not reveal the symmetry of the superconducting order parameter.

Complementary information on the gap structure is provided by Andreev reflections. This becomes most evident at zero bias $V=0$, where the differential conductance is fully controlled by Andreev reflections, $|\alpha_{p,h}|=1$ and hence $|r_{ph}|^2+|r_p|^2=1$. In the corresponding limit $|\varepsilon|\to 0$, the Andreev amplitudes, Eq.~(\ref{alpha}) are $\alpha_p(\mathbf{k})=-\alpha_h^*(\mathbf{k})=-i\Delta(\mathbf{k})/|\Delta(\mathbf{k})|$. We can then evaluate  Eqs.~(\ref{dIdV}) and (\ref{rAfinal})  for arbitrary junction conductance $G_n$ and obtain
\begin{eqnarray}
&&G(V=0,\mathbf{r}_0)=2G_Q|r_{ph}(\varepsilon=0,\mathbf{r}_0)|^2\,, \label{GAzerobias}\\
&&r_{ph}=\frac{(2/i)(1-|s_0|^2)\langle
    |u_{\mathbf{k}}({\bf r}_0)|^2\Delta(\mathbf{k})/|\Delta(\mathbf{k})|\rangle_0}
    {|1+s_0|^2+|1-s_0|^2\left|\langle
    |u_{\mathbf{k}}({\bf r}_0)|^2\Delta(\mathbf{k})/|\Delta(\mathbf{k})|\rangle_0\right|^2}.\nonumber
\end{eqnarray}
This expression shows that the zero-bias conductance is sensitive to the interplay of the symmetries of the Bloch functions and the superconducting gap. Since the symmetry of the Bloch function varies with the tip position $\mathbf{r}_0$, it provides a powerful tool to extract the gap structure.

If $\mathbf{r}_0$ is invariant under the lattice point-symmetry group, then $u_{\mathbf{k}}({\bf r}_0)$ as a function of $\mathbf{k}$ belongs to an irreducible representation of the point group. Assuming that the only degeneracy of the Bloch states at the Fermi energy is associated with TRS,  $u_{\mathbf{k}}({\bf r}_0)$ belongs to a one-dimensional representation, {\sl i.e.}, $u_{\mathbf{k}}({\bf r}_0)$ acquires only a phase factor and $|u_{\mathbf{k}}({\bf r}_0)|^2$ is invariant under point-group operations. In contrast, there is no corresponding symmetry requirement when $\mathbf{r}_0$ is a generic point within the unit cell. Now consider the symmetry of $\Delta(\mathbf{k})/|\Delta(\mathbf{k})|$, entering into Eq.~(\ref{GAzerobias}). First we note that $\sqrt{\xi^2({\bf k})+|\Delta(\mathbf{k})|^2}$ is an eigenvalue of the Bogoliubov-de-Gennes (BdG) Hamiltonian. If $\Delta(\mathbf{k})$ does not break the lattice symmetry, then the eigenvalues of the BdG Hamiltonian, as well as $\xi(\bf{k})$ are invariant under point group transformations. Thus, $|\Delta(\mathbf{k})|$ belongs to the trivial representation~\cite{Geier-Brouwer-Trifunovic:2019}, while $\Delta(\mathbf{k})/|\Delta(\mathbf{k})|$ together with $\Delta(\mathbf{k})$ belongs to some representation of the lattice point group. If that representation is trivial (as for $s$-wave superconductivity), then $\Delta(\mathbf{k})/|\Delta(\mathbf{k})|$ is independent of $\mathbf{k}$ and Eq.~(\ref{GAzerobias}) reproduces the conventional result~\cite{BTK:1982} $|r_{ph}|=(1-|s_0|^2)/(1+|s_0|^2)$, even if $\mathbf{r}_0$ is not a lattice symmetry point. With increasing tunneling strength, $G(V=0)$ varies from $\sim G_n^2/G_Q$  at $s_0\to1$ to the saturation value $2G_Q$ at $s_0=0$.

If $\Delta(\mathbf{k})$ belongs to a nontrivial representation of the point group, then at a high-symmetry point the V shape of $dI/dV$ with $G(V=0,\mathbf{r}_0)=0$ persists for any $G_n$; see Fig.~\ref{fig:G}(b), but $G(V=0,\mathbf{r}_0)$ is finite at a generic $\mathbf{r}_0$. Lastly, if $\Delta(\mathbf{k})$ breaks the lattice symmetry, one expects a nonzero, position-dependent $G(V=0,\mathbf{r}_0)$; depending on details, $G(0)$ may or may not reach the saturation value $2G_Q$, see also Figs.~\ref{fig:G}(c) and \ref{fig:G}(e). One may understand these results pictorially; see Fig.~\ref{fig:STM-Andreev}. The total Andreev-reflection amplitude is a superposition of partial ones coming from the different directions ${\bf k}/|\mathbf{k}|$. Each partial amplitude carries a phase, governed by the gap an injected particle ``sees" in the given direction. For a real-valued and symmetric nodal gap, the negative and positive contributions to the sum cancel each other. The presence of Bloch functions may lift the cancellation if their symmetry is different from that of the gap, or if the tunneling point is away from a high-symmetry point.

The conductance $G(V)$ depends strongly on the strength $G_n$ of the tunneling contact; see Fig.~\ref{fig:G}. Focusing on the strong-tunneling limit of $s_0=0$ ({\sl i.e.}, $G_n=G_Q$), we can analytically extract the asymptotes of $G(V)$ for $V\to 0$ and $V\to \Delta$~\cite{SM}.

We start with the $V\to 0$ asymptote. For a real-valued gap without nodal points (TRS is preserved, but spatial symmetry may be broken), we find $G(V)=2G_Q[1-\gamma^{\phantom C}_R (eV/\Delta)^4]$ to  leading nontrivial order in $eV/\Delta$. The  coefficient $\gamma^{\phantom C}_R>0$ depends on  details of the gap structure as well as $\mathbf{r}_0$. For isotropic gaps, $\gamma^{\phantom C}_R=0$ at any $\mathbf{r}_0$ and Eq.~(\ref{dIdV}) is identical to known results in a one-dimensional geometry~\cite{BTK:1982}. A real-valued gap with nodal points leads to $G(V)=G(0)+G_Q\gamma^{\phantom C}_V |eV|/\Delta$ with the sign of the coefficient $\gamma^{\phantom C}_V$ depending on details of the gap and the tip position; for gaps respecting the lattice symmetry and $\mathbf{r}_0$ located at a symmetry point, $G(0)=0$ and $\gamma^{\phantom C}_V>0$, see Fig.~\ref{fig:G}(b). If the gap is complex valued and nodeless (broken TRS), but does not break the point- group symmetry (as in a ${\rm d}_{x^2-y^2}+i{\rm d}_{xy}$ superconductor), we find $G(V)=0$ in the entire interval $|eV|<\mbox{min}{\left\{\Delta(\mathbf{k})\right\}}$ for tunneling at a symmetry point; see Fig.~\ref{fig:G}(d). Away from symmetry points, $G(V)=G(0) -G_{Q} \gamma_{C} \left|eV/\Delta\right|^2$ with model-dependent values of $G(0)$ and $\gamma_{C}$. If the point-group symmetry is broken in addition to TRS (as in  noncollinear ${\rm A}_2+{\rm E}_1+i{\rm E}_2$ states~\cite{Chichinadze-Chubukov:2020}), then $G(V)=G(0) -G_Q\gamma_C |eV/\Delta|^2$ with $G(0)<2G_Q$ regardless of tip position; see Fig.~\ref{fig:G}(e). The coefficients $\gamma^{\phantom C}_C$ in the last two asymptotes depend on the specific gap structure.

Extrema $\Delta_{\rm extr}$ in $|\Delta ({\mathbf{k}})|$ lead to van Hove singularities in the tunneling density of states, which appear as ``coherence peaks'' $\propto\ln({\Delta_{\rm extr}/|\Delta_{\rm extr}-eV|})$ in the tunneling conductance at $G_n\ll G_Q$. At stronger tunneling, the peaks turn into singular minima of the form $A+B\ln^{-1}({\Delta_{\rm extr}/|\Delta_{\rm extr}-eV|})$ analogous to Fano resonances [Fig.~\ref{fig:G}(b)]. This structure becomes most prominent at full transmission ($s_0=0$), where $G(V)$ may vanish at the singularity, see, {\sl e.g.}, Figs.~\ref{fig:G}(b) and \ref{fig:G}(e)~\cite{SM}.

{\it Discussion and Summary.}
Our theory summarized in Eqs.~(\ref{aph2})--(\ref{dIdV}) describes the differential conductance $G(V)$ in an STS setting for a 2D superconductor  at arbitrary junction transmission as well as arbitrary symmetries of the order parameter and Bloch functions. The zero-bias results are expressed, in an intuitive way, by Eq.~(\ref{GAzerobias}). We used the theory to perform a symmetry analysis of the conductance and make specific predictions for tunneling both at and away from high-symmetry points of the lattice; see Fig.~\ref{fig:G} and Table~S1~\cite{SM} for further details.

Moir\'{e} materials such as TBG have a Fermi wavelength that is much larger than that of the metallic tip. Thus the single-channel-contact approximation is adequate unless the normal conductance $G_n$ exceeds $G_Q$, indicating a substantial increase in a contact area. As long as the contact preserves its single-channel nature, the observation of a zero-bias conductance maximum at strong tunneling along with a prominent V-shaped conductance at weak tunneling, as reported in~\cite{Oh-Yazdani:2021}, is incompatible with a nodal gap respecting the lattice point symmetry. Indeed, in the latter case the low-bias behavior of $G(V)$ is linear at any tunneling strength;  see, {\sl e.g.}, Fig.~\ref{fig:G}(b). The experimental data~\cite{Oh-Yazdani:2021} for filling factors between $-2$ and $-3$ may be consistent with a strongly anisotropic gap with a small $\Delta_{\rm min}$, as exemplified in Fig.~\ref{fig:G}(c). However, while the superconducting gap symmetry of TBG is unknown, the required fine-tuning (e.g., between the strengths of $s$- and $d$-wave orders) would hardly persist over the  entire filling-factor range~\cite{senthil-privatecomm}. A possible resolution~\cite{Yazdani-privatecomm} of this conundrum is provided by the data in Fig.~S6 of Ref.~\cite{Oh-Yazdani:2021}. There, the differential conductance is V-shaped as long as $G(V)$ remains below the maximal single-channel conductance $2G_Q$ for Andreev reflection. The V-shaped traces evolve into a zero-bias maximum only upon further increasing the junction conductance, where the tip may have developed a contact area of the order of the Moir\'{e} period and thus created a multichannel junction~\cite{Yazdani-privatecomm}.

The differential conductance $G(V)$ in the STM setting was also recently obtained numerically in Ref.~\cite{Lake-Senthil:2022}, using the tunneling Hamiltonian approach. For $d$- or $p$-wave superconductivity of TBG, the V-shaped dependence and the absence of a zero-bias peak persist for all tunneling strengths $\mathsf{t}$. Our theory, applied under the same conditions, is consistent with the conclusions of Ref.~\cite{Lake-Senthil:2022}, but also more nuanced. First, including the Bloch functions accounts for the dependence of $G(V)$ on the point of tunneling. In particular, $G(V=0)$ may be nonzero, even if the gap does not break the lattice symmetry. Second, our fully analytical solution based on scattering theory overcomes limitations of the tunneling Hamiltonian. At partial transmission, $G(V)$ depends not only on $\mathsf{t}=\sqrt{1-|s_0|^2}$, but also on the \emph{phase} of the scattering amplitude $s_0$. Accounting for the phase is important even at the qualitative level, affecting, {\sl e.g.}, the $V\to -V$ symmetry of $G(V)$~\cite{SM}. Our analytical solution also exposes the low-bias behavior of $G(V)$ and the emergence of a Fano resonance at stronger tunneling; see Fig.~\ref{fig:G}(b).

While we made several simplifying assumptions, our method applies more generally and allows for various extensions. For example, we assumed that, in the absence of tunneling, the tip does not create a scattering potential within the 2D material, {\sl i.e.}, $s_0(\varepsilon)=1$. Such a potential is readily incorporated through a scattering phase in $s_0$, leading to subgap resonances. Thus, our work provides a flexible and powerful framework to analyze future STM experiments aimed at revealing and analyzing the structure of the superconducting gap in TBG and other novel 2D superconductors.

\begin{acknowledgments}
This work was motivated by a discussion with Ali Yazdani at the Aspen Center for Physics supported by NSF Grant No.~PHY-1607611. We are grateful to Piet Brouwer, Katharina Franke, and Vlad Kurilovich for illuminating remarks, and T.~Senthil and Ali Yazdani for helpful comments. This work is supported by NSF Grant No.~DMR-2002275 (L.I.G.), Deutsche Forschungsgemeinschaft through CRC 183 (Mercator fellowship, L.I.G.; project C02, F.v.O.) and a joint ANR-DFG project (TWISTGRAPH, F.v.O.). P.O.S. acknowledges support through the Yale Prize Postdoctoral Fellowship in Condensed Matter Theory.
\end{acknowledgments}

\vspace{-0.7cm}

\bibliography{library-short}

\end{document}


\title{Supplemental Material\\
\mbox{Andreev reflection in scanning tunneling spectroscopy of unconventional superconductors}}

\author{P.~O.~Sukhachov}
\email{pavlo.sukhachov@yale.edu}
\affiliation{Department of Physics, Yale University, New Haven, Connecticut 06520, USA}

\author{Felix von Oppen}
\affiliation{\mbox{Dahlem Center for Complex Quantum Systems and Fachbereich Physik, Freie Universit\"{a}t Berlin, 14195 Berlin, Germany}}

\author{L.~I.~Glazman}
\affiliation{Department of Physics, Yale University, New Haven, Connecticut 06520, USA}

\maketitle
\tableofcontents

\section{S I. Details of the derivation of Eqs.~(12) and (13) in the main text}
\label{sec:app-S1}

In this Section, we solve the integral equation (8) in the main text, which we reproduce here for convenience,
\begin{equation}
\label{app-S1-B-M-int}
\left(\hat{I} -\hat{L}\right)M(\mathbf{k}) =  \alpha_p(\mathbf{k}) u_{\mathbf{k}}(\mathbf{r}_0),
\end{equation}
and calculate the scattering amplitudes of particles and holes in the STM setup.

Since the operator $\hat{L} =  \alpha_p(\mathbf{k}) \left[\hat{I}-\left(1-s_0\right)\hat{P}\right] \alpha_h(\mathbf{k}) \left[\hat{I}-\left(1-s_0^*\right)\hat{P}\right]$ defined in Eq.~(6) in the main text has a separable kernel, the integral equation (\ref{app-S1-B-M-int}) can be brought to a set of algebraic equations; see, {\sl e.g.}, Ref.~\cite{Morse-Feshbach:book}. Indeed, the action of the operator $\hat{L}$ on $M(\mathbf{k})$ reads
\begin{eqnarray}
\label{app-S1-B-M-int-expl}
\hat{L} M(\mathbf{k}) &=&
\alpha_{p}(\mathbf{k})\alpha_{h}(\mathbf{k}) M(\mathbf{k})
-\left(1-s_{0}^{*}\right) \alpha_{p}(\mathbf{k}) \alpha_{h}(\mathbf{k})u_{\mathbf{k}}(\mathbf{r}_0) M_1
- \left(1-s_{0}\right) \alpha_{p}(\mathbf{k}) u_{\mathbf{k}}(\mathbf{r}_0) M_2 \nonumber\\
&+& \left|1-s_{0}\right|^2 \alpha_{p}(\mathbf{k}) u_{\mathbf{k}}(\mathbf{r}_0) \left\langle |u_{\mathbf{k}}(\mathbf{r}_0)|^2 \alpha_{h}(\mathbf{k}) \right\rangle_{\varepsilon} M_1,
\end{eqnarray}
where
\begin{equation}
\label{app-S1-B-M1-M2}
M_1 = \left\langle u_{\mathbf{k}}^{*}(\mathbf{r}_0) M(\mathbf{k})\right\rangle_{\varepsilon} \quad \mbox{and} \quad M_2 =\left\langle u_{\mathbf{k}}^{*}(\mathbf{r}_0) \alpha_h(\mathbf{k}) M(\mathbf{k})\right\rangle_{\varepsilon}.
\end{equation}
By substituting Eq.~(\ref{app-S1-B-M-int-expl}) into Eq.~(\ref{app-S1-B-M-int}), we find the following expression for $M(\mathbf{k})$ in terms of $M_1$ and $M_2$:
\begin{eqnarray}
\label{app-S1-B-M-eq}
M(\mathbf{k}) &=& \frac{u_{\mathbf{k}}(\mathbf{r}_0) \alpha_p(\mathbf{k})}{1-\alpha_{p}(\mathbf{k})\alpha_h(\mathbf{k})}
+ \left|1-s_{0}\right|^2 \frac{u_{\mathbf{k}}(\mathbf{r}_0) \alpha_p(\mathbf{k})}{1-\alpha_p(\mathbf{k})\alpha_h(\mathbf{k})} \left\langle |u_{\mathbf{k}}(\mathbf{r}_0)|^2 \alpha_{h}(\mathbf{k}) \right\rangle_{\varepsilon} M_1
-\left(1-s_{0}^{*}\right) \frac{u_{\mathbf{k}}(\mathbf{r}_0) \alpha_p(\mathbf{k}) \alpha_h(\mathbf{k})}{1-\alpha_p(\mathbf{k}) \alpha_h(\mathbf{k})} M_1
\nonumber\\
&-& \left(1-s_{0}\right) \frac{u_{\mathbf{k}}(\mathbf{r}_0) \alpha_p(\mathbf{k})}{1-\alpha_p(\mathbf{k})\alpha_h(\mathbf{k})} M_2.
\end{eqnarray}
Using this result in Eq.~(\ref{app-S1-B-M1-M2}), we obtain the set of algebraic equations for $M_1$ and $M_2$:
\begin{eqnarray}
\label{app-S1-B-M-eq-1}
&&\left[1 -\left|1-s_{0}\right|^2 a_{p} \left\langle |u_{\mathbf{k}}(\mathbf{r}_0)|^2 \alpha_{h}(\mathbf{k}) \right\rangle_{\varepsilon} +\left(1-s_{0}^{*}\right) a_{ph}\right]M_1
+\left(1-s_{0}\right) a_{p} M_2 =a_p,\\
\label{app-S1-B-M-eq-2}
&&\left[\left(1-s_{0}^{*}\right) a_{phh} -\left|1-s_{0}\right|^2 a_{ph} \left\langle |u_{\mathbf{k}}(\mathbf{r}_0)|^2 \alpha_{h}(\mathbf{k}) \right\rangle_{\varepsilon} \right]M_1
+\left[1 +\left(1-s_{0}\right) a_{ph}\right]M_2=a_{ph}.
\end{eqnarray}
Here, we used the following shorthand notations:
\begin{align}
\label{app-S1-B-ap-aph-def}
a_{p} &= \left\langle |u_{\mathbf{k}}(\mathbf{r}_0)|^2 \frac{\alpha_{p}(\mathbf{k})}{1-\alpha_{p}(\mathbf{k})\alpha_{h}(\mathbf{k})} \right\rangle_{\varepsilon}, &\quad a_{ph} &= \left\langle |u_{\mathbf{k}}(\mathbf{r}_0)|^2 \frac{\alpha_{p}(\mathbf{k})\alpha_{h}(\mathbf{k})}{1-\alpha_{p}(\mathbf{k})\alpha_{h}(\mathbf{k})} \right\rangle_{\varepsilon},\\
\label{app-S1-B-ah-aphh-def}
a_{h} &= \left\langle |u_{\mathbf{k}}(\mathbf{r}_0)|^2 \frac{\alpha_{h}(\mathbf{k})}{1-\alpha_{p}(\mathbf{k})\alpha_{h}(\mathbf{k})} \right\rangle_{\varepsilon},  &\quad a_{phh} &= \left\langle |u_{\mathbf{k}}(\mathbf{r}_0)|^2 \frac{\alpha_{p}(\mathbf{k})\alpha_{h}^2(\mathbf{k})}{1-\alpha_{p}(\mathbf{k})\alpha_{h}(\mathbf{k})} \right \rangle_{\varepsilon}.
\end{align}

By solving the system of algebraic equations (\ref{app-S1-B-M-eq-1}) and (\ref{app-S1-B-M-eq-2}), we obtain
\begin{eqnarray}
\label{app-S1-B-M-eq-sol-1}
M_1 &=& \frac{a_{p}}{1 + \left(2-s_{0}-s_{0}^{*}\right)a_{ph} +\left|1-s_{0}\right|^2 \left(a_{ph}^2 -a_{p} a_{h}\right)},\\
\label{app-S1-B-M-eq-sol-2}
M_2 &=& \frac{a_{ph} +\left(1-s_0^{*}\right)\left[a_{ph}^2 -a_p\left(a_{h} -\left\langle |u_{\mathbf{k}}(\mathbf{r}_0)|^2 \alpha_{h}(\mathbf{k}) \right\rangle_{\varepsilon}\right) \right]}{1 + \left(2-s_{0}-s_{0}^{*}\right)a_{ph} +\left|1-s_{0}\right|^2 \left(a_{ph}^2 -a_{p} a_{h}\right)}.
\end{eqnarray}
The above equations allow us to derive the Andreev scattering amplitude
\begin{equation}
\label{app-S1-B-rph-fin}
r_{ph} =|t|^2\langle u_{\mathbf{k}}^{*}({\bf r}_0){M}(\mathbf{k})\rangle^{\phantom\dagger}_\varepsilon = |t|^2 M_1
= \frac{|t|^2a_p}{1 + \left(2-s_{0}-s_{0}^{*}\right)a_{ph} +\left|1-s_{0}\right|^2 \left(a^2_{ph} -a_p a_h\right)},
\end{equation}
which is given in Eq.~(12) in the main text. In the case of impinging holes, the corresponding amplitude $r_{\rm hp}$ is obtained by replacing $a_p\leftrightarrow a_h$ in Eq.~(\ref{app-S1-B-rph-fin}).

Let us turn our attention to the normal reflection amplitude $r_{p}$. By including an additional half-a-cycle, in which a hole is scattered off the tip and is retroreflected as a particle, we obtain
\begin{eqnarray}
\label{app-S1-B-r-pp}
r_{p} &=& s_0^{\prime} + t \left\langle u_{\mathbf{k}}^{*}({\bf r}_0) \alpha_h(\mathbf{k}) \left[\hat{I}-\left(1-s_0^*\right)\hat{P}\right] M(\mathbf{k}) \right\rangle_{\varepsilon} = s_0^{\prime} + t^2\left[M_2 -\left(1-s_0^*\right)\left\langle |u_{\mathbf{k}}(\mathbf{r}_0)|^2 \alpha_h(\mathbf{k})\right\rangle_{\varepsilon} M_1\right] \nonumber\\
&=& s_0^{\prime} + \frac{t^2\left[a_{ph} +\left(1-s_{0}^{*}\right) \left(a_{ph}^2 -a_{p}a_{h}\right)\right]}{1 + \left(2-s_{0}-s_{0}^{*}\right)a_{ph} +\left|1-s_{0}\right|^2 \left(a_{ph}^2 -a_{p}a_{h}\right)}.
\end{eqnarray}
The expression in the last line corresponds to Eq.~(13) in the main text. In the case of impinging holes, the amplitude $r_{h}$ is obtained  by replacing $s_0 \leftrightarrow s_0^{*}$ and $t \leftrightarrow t^{*}$ in Eq.~(\ref{app-S1-B-r-pp}).

\section{S II. Differential conductance: particle-hole symmetry and strong tunneling }
\label{sec:app-S2}

\subsection{S II.A Explicit form of the conductance bias dependence}
\label{sec:app-S2-A}

We combine the presented in the main text Eq.~(14) and its extension to negative biases into a single equation for the differential conductance
\begin{equation}
\label{app-S1-B-Gtot-def}
\frac{dI(V,\mathbf{r}_0)}{dV} =G(V,\mathbf{r}_0) = G_Q \left\{1+\left[\left|r_{ph}(|eV|,\mathbf{r}_0)\right|^2-\left|r_{p}(|eV|,\mathbf{r}_0)\right|^2\right]\theta(V)
+\left[\left|r_{hp}(|eV|,\mathbf{r}_0)\right|^2-\left|r_{h}(|eV|,\mathbf{r}_0)\right|^2\right]\theta(-V)\right\}.
\end{equation}
Here, $G_{Q}=e^2/(\pi \hbar)$ is the conductance quantum, $\theta(V)$ is the step function, the reflection amplitudes are given in Eqs.~(\ref{app-S1-B-rph-fin}) and (\ref{app-S1-B-r-pp}), and, as in the main text, temperature is set to zero.

We find it convenient to separate the parts corresponding to the subgap $|\varepsilon|\leq \mbox{min}{\left\{|\Delta(\mathbf{k})|\right\}}\equiv \Delta_{\rm min}$ and above-the-gap $|\varepsilon|\geq \mbox{max}{\left\{|\Delta(\mathbf{k})|\right\}} \equiv \Delta_{\rm max}$ energies in Eqs.~(\ref{app-S1-B-ap-aph-def}) and (\ref{app-S1-B-ah-aphh-def}),
\begin{equation}
\label{app-S2-ap}
a_{p} = \frac{1}{2i} \left(B_{<} +i \, B_{>}\right), \quad a_{ph} =\frac{1}{2i} \left(A_{<} +i \, A_{>}\right) - \frac{1}{2}, \quad a_{h} = \frac{1}{2i} \left(B_{<}^{*} +i \, B_{>}^{*}\right).
\end{equation}
The explicit form of the coefficients $A_{<,>}$ and $B_{<,>}$ is
\begin{align}
\label{app-S1-B-BRe-def}
B_{<} &= \left\langle \left|u_{\mathbf{k}}(\mathbf{r}_0)\right|^2 \frac{\Delta^{*}(\mathbf{k})\,\theta\left(|\Delta(\mathbf{k})|^2-\varepsilon^2\right)}{\sqrt{|\Delta(\mathbf{k})|^2-\varepsilon^2}} \right\rangle_{\varepsilon}, &\quad
B_{>} &= \left\langle \left|u_{\mathbf{k}}(\mathbf{r}_0)\right|^2 \frac{\Delta^{*}(\mathbf{k}) \,\theta\left(\varepsilon^2-|\Delta(\mathbf{k})|^2\right)}{\sqrt{\varepsilon^2-|\Delta(\mathbf{k})|^2}} \right\rangle_{\varepsilon},\\
\label{app-S1-B-ARe-def}
A_{<} &= \varepsilon \,\left\langle \left|u_{\mathbf{k}}(\mathbf{r}_0)\right|^2 \frac{\theta\left(|\Delta(\mathbf{k})|^2-\varepsilon^2\right)}{\sqrt{|\Delta(\mathbf{k})|^2-\varepsilon^2}} \right\rangle_{\varepsilon}, & \quad
A_{>} &= \varepsilon \,\left\langle \left|u_{\mathbf{k}}(\mathbf{r}_0)\right|^2 \frac{\theta\left(\varepsilon^2-|\Delta(\mathbf{k})|^2\right)}{\sqrt{\varepsilon^2-|\Delta(\mathbf{k})|^2}} \right\rangle_{\varepsilon},
\end{align}
where we used the retroreflection amplitudes $\alpha_{p,h}$ given in Eq.~(3) in the main text. We also introduce the following shorthand notations:
\begin{equation}
\label{app-S1-C-CRe-def}
C_{1} = |B_{<}|^2-|B_{>}|^2 -A_{<}^2+A_{>}^2 -1, \quad \quad C_{2} = 2\,\mbox{Re}{\left\{B_{<}^{*} \,B_{>}\right\}} -2A_{<} \,A_{>}.
\end{equation}

Equation~(\ref{app-S2-ap}) allows us to rewrite the absolute values of the reflection amplitudes (\ref{app-S1-B-rph-fin}) and (\ref{app-S1-B-r-pp}) in a compact form,
\begin{align}
\label{app-S1-C-rp2}
\left|r_{p}\right|^2 &= \left|-s_0^{*}+|t|^2\frac{\mathcal{N}}{\mathcal{D}}\right|^2, &\quad
\left|r_{h}\right|^2 &= \left|-s_0+|t|^2\frac{\mathcal{N}}{\mathcal{D}}\Big|_{s_0^{*}\to s_0}\right|^2,\\
\label{app-S1-C-rph2}
\left|r_{ph}\right|^2 &= |t|^4 \frac{|B_{<}|^2+|B_{>}|^2 +2\,\mbox{Im}{\left\{B_{<}\, B_{>}^{*}\right\}} }{4|\mathcal{D}|^2}, &\quad  \left|r_{hp}\right|^2 &= |t|^4 \frac{|B_{<}|^2+|B_{>}|^2 -2\,\mbox{Im}{\left\{B_{<}\, B_{>}^{*}\right\}} }{4|\mathcal{D}|^2}.
\end{align}
Here, $s_0^{\prime} = -s_0^{*} t/t^{*}$, which follows from the unitarity of the scattering matrix, was used in Eq.~(\ref{app-S1-C-rp2}). The terms $\mathcal{D}$ and $\mathcal{N}$ in the above equations are
\begin{eqnarray}
\label{app-S1-C-D}
\mathcal{D}\! &=&\! 1 + \left(2-s_{0}-s_{0}^{*}\right)a_{ph} +\left|1-s_{0}\right|^2 \left(a_{ph}^2 -a_{p}a_{h}\right)
= \frac{1+|s_0|^2}{2} -i \frac{|t|^2}{2} \left(A_{<} +i \,A_{>}\right) +\frac{|1-s_0|^2}{4} \left(C_{1} +i \,C_{2}\right),\\
\label{app-S1-C-N}
\mathcal{N}\! &=&\! a_{ph} +\left(1-s_{0}^{*}\right) \left(a_{ph}^2 -a_{p}a_{h}\right)
=-\frac{s_0^{*}}{2} -i\frac{s_0^{*}}{2} \left(A_{<} +i \,A_{>}\right) +\frac{1-s_0^{*}}{4} \left(C_{1} +i \,C_{2}\right),
\end{eqnarray}
where we used Eqs.~(\ref{app-S2-ap}) and (\ref{app-S1-C-CRe-def}). The obtained in this Section expressions allow us to calculate the differential conductance at any value of the scattering matrix element $s_0$.

\subsection{S II.B Particle-hole symmetry constraint}
\label{sec:app-S2-B}

Regardless of the presence of time-reversal symmetry (TRS), the differential conductance is symmetric in bias, $G(V)=G(-V)$, at $|eV|< \Delta_{\rm min}$ and $|eV|>\Delta_{\rm max}$. On the other hand,
the conductance may develop asymmetry in the intermediate interval, $\Delta_{\rm min} < |eV| < \Delta_{\rm max}$.

In the presence of TRS, the only possible source of asymmetry is the difference between $|r_p|$ and $|r_h|$. First, we notice that the denominator in Eq.~(\ref{app-S1-B-r-pp}) is particle-hole ({\it ph}) symmetric, see also Eq.~(\ref{app-S1-C-D}) for the explicit form of the denominator $\mathcal{D}$. Terms $a_{ph}$ and $a_{ph}^2-a_pa_h$ are also {\it ph}-symmetric. We introduce notations, $X_{ph}=\left(a_{ph}+a_{ph}^2-a_pa_h\right)/\mathcal{D} =Y_{ph}+ a_{ph}/\mathcal{D}$ and $Y_{ph}=\left(a_{ph}^2-a_pa_h\right)/\mathcal{D}$, emphasizing the {\it ph} symmetry: $X_{ph}=X_{hp}$, $Y_{ph}=Y_{hp}$. Using these notations and the $S$-matrix unitarity condition $s_0^{\prime} = -s_0^{*} t/t^{*}$, we re-write the amplitudes $r_p$ and $r_h$ as
\begin{equation}
\label{rprh1}
r_p = -\frac{t}{t^{*}} \left[ s_0^{*} \left(1+|t|^2Y_{ph}\right) -|t|^2X_{ph}\right]\,,\quad
r_h = -\frac{t^*}{t} \left[ s_0 \left(1+|t|^2Y_{ph}\right) -|t|^2X_{ph}\right]\,,
\end{equation}
and find after simple algebra:
\begin{equation}
\label{app-S2-C-rp2-rh2}
|r_p|^2 -|r_h|^2 = -4|t|^2 \,\mbox{Im}{\left\{s_0\right\}}\, \mbox{Im}{\left\{\left(1+|t|^2Y_{ph}\right)X_{ph}^{*}\right\}}.
\end{equation}
Using here the explicit form of $\mathcal{D}$ given in Eq.~(\ref{app-S1-C-D}) and Eq.~(\ref{app-S2-ap}), we find
\begin{equation}
\label{rprh2}
|r_p|^2 -|r_h|^2 \propto
\mbox{Im}\{s_0\} \mbox{Im}\{\left(a_{ph}+a_{ph}^2-a_pa_h\right)^{*}\}
=\frac{1}{2}\mbox{Im}\{s_0\}\,
\left(A_{<} \,A_{>}-\mbox{Re}{\left\{B_{<}^{*} \,B_{>}\right\}}\right).
\end{equation}
The last factor here can differ from zero only if $\Delta_{\rm min}<|eV|<\Delta_{\rm max}$; see Eqs.~(\ref{app-S1-B-BRe-def}) and (\ref{app-S1-B-ARe-def}) for the definitions of $B_{<,>}$ and $A_{<,>}$.
Furthermore, the asymmetry between reflection amplitudes (\ref{rprh2}) relies on the presence of the phase of the scattering amplitude, $\mbox{Im}\{s_0\} \neq0$.
Lastly, $|r_{ph}|=|r_{hp}|$, as $a_p=a_h$ is enforced by TRS.

The origin of the symmetry violation uncovered in Eq.~(\ref{rprh2}) is the interference between the processes of conversion of an incoming electron into a quasiparticle propagating in the superconductor. One of the processes is a direct conversion into a quasiparticle propagating in some direction ${\bf k}$, whereas the other one is a conversion into the same state upon completion of an integer number of Andreev reflection cycles involving other directions.

In the absence of TRS, the relation $|r_{ph}|=|r_{hp}|$ still holds outside the interval $\left(\Delta_{\rm min},\Delta_{\rm max}\right)$ as $|a_p|=|a_h|$ there. The latter two relations may be violated at biases within the interval $\Delta_{\rm min}<|eV|<\Delta_{\rm max}$, see Eqs.~(\ref{app-S2-ap}) and (\ref{app-S1-B-BRe-def}).

\subsection{S II.C Simplified form of \texorpdfstring{$G(V)$}{G(V)} in the strong tunneling regime (\texorpdfstring{$s_0=0$}{s0=0})}
\label{sec:app-S2-C}

In the case of strong tunneling $s_0=0$, the absolute values of $\mathcal{D}$ and $\mathcal{N}$ in Eqs.~(\ref{app-S1-C-D}) and (\ref{app-S1-C-N}) are
\begin{eqnarray}
\label{app-S2-2-D2}
|\mathcal{D}|^2 \!&=&\! \frac{1}{4} \left[1 +A_{<}^2 +A_{>}^2 -\left(A_{<}C_{2} -A_{>}C_{1}\right) +\frac{1}{4} \left(C_{1}^2 +C_{2}^2\right) + C_{1} +2 A_{>} \right] \nonumber\\
&=&\frac{1}{16} \left\{\left[\left(1+A_{>}\right)^2 -\left(A_{<}^2 +|B_{>}|^2 -|B_{<}|^2\right) \right]^2
+4\left[A_{<}\left(1+A_{>}\right) -\mbox{Re}{\left\{B_{>}^{*}\,B_{<}\right\}}\right]^2\right\},\\
\label{app-S2-2-N2}
|\mathcal{N}|^2 \!&=&\! \frac{1}{16} \left(C_{1}^2 +C_{2}^2\right) = |\mathcal{D}|^2 -\frac{1}{4}\left[A_{>} \left(1 +A_{>}^2+A_{<}^2 +|B_{<}|^2 -|B_{>}|^2\right)
+2A_{>}^2 +|B_{<}|^2-|B_{>}|^2  -2A_{<} \mbox{Re}{\left\{B_{<}^{*}\, B_{>}\right\}}
\right].\nonumber\\
\end{eqnarray}
By using Eqs.~(\ref{app-S1-C-rp2}) and (\ref{app-S1-C-rph2}) at $s_0=0$ with the expressions (\ref{app-S2-2-D2}) and (\ref{app-S2-2-N2}) in Eq.~(\ref{app-S1-B-Gtot-def}), we obtain
\begin{equation}
\label{app-S3-A-3-G-def}
G(V) =4G_Q \frac{A_{>}\left[\left(1+A_{>}\right)^2 +A_{<}^2 +|B_{<}|^2 -|B_{>}|^2\right] +2|B_{<}|^2 -2A_{<}\,\mbox{Re}{\left\{B_{<}\,B_{>}^{*}\right\}} +2\sign{V}\,\mbox{Im}{\left\{B_{<}\,B_{>}^{*}\right\}}}{\left[\left(1+A_{>}\right)^2 -\left(A_{<}^2 +|B_{>}|^2 -|B_{<}|^2\right)\right]^2 +4\left[A_{<}\left(1+A_{>}\right) -\mbox{Re}{\left\{B_{<}\,B_{>}^{*}\right\}}\right]^2},
\end{equation}
where the conductance depends on $|eV|=\varepsilon$ and the tip position $\mathbf{r}_0$ via the coefficients $B_{<,>}$ and $A_{<,>}$ defined in Eqs.~(\ref{app-S1-B-BRe-def}) and (\ref{app-S1-B-ARe-def}).

Equation~(\ref{app-S3-A-3-G-def}) can be simplified in the case of the subgap, $|eV|< \Delta_{\rm min}$, or above-the-gap, $|eV|> \Delta_{\rm max}$, energies. We have $A_{>}=B_{>}=0$ in the former case and $A_{<}=B_{<}=0$ in the latter one. The conductance in these cases reads
\begin{eqnarray}
\label{app-S2-2-Gtot-sub}
|eV|<\Delta_{\rm min}: \quad G(V) &=& 8G_Q \frac{|B_{<}|^2}{\left[1 -\left(A_{<}^2-|B_{<}|^2\right)\right]^2 +4 A_{<}^2},\\
\label{app-S2-2-Gtot-super}
|eV|>\Delta_{\rm max}: \quad G(V) &=& 4G_Q \frac{A_{>}}{\left(1+A_{>}\right)^2 -|B_{>}|^2}.
\end{eqnarray}

\section{S III. Conductance bias dependence and numerical results at \texorpdfstring{$s_0=0$}{s0=0}}
\label{sec:app-S3}

In this Section, we discuss the dependence of the conductance on voltage bias to illustrate the general symmetry arguments presented in the main text. We focus on the regime of the strong tunneling ($s_0=0$) discussed in Sec.~S~II.C, and consider three types of the gap: (i) nodeless TRS-preserving (real-valued) gap, (ii) nodal gap, and (iii) nodeless TRS-breaking (complex-valued) gap. For an arbitrary structure of the gap, the Bloch function, and the scattering matrix element $s_0$, the conductance can be calculated numerically via the expressions in Sec.~S~II.A. However, even in a general case, we still can, using analytical means, extract information about the low-bias behavior of the conductance as well as elucidate the peculiarities of the $G(V)$ dependence associated with the van Hove singularities in the Bogoliubov quasiparticle spectrum. We summarize the bias profiles of the conductance for a few characteristic gap structures in Tab.~\ref{tab:crossover-table}. The details of the derivation and the explicit form of the coefficients can be found in Secs.~S~III.A and S~III.B for small energies and near the van Hove singularities, respectively.

\begin{table*}
\begin{tabular}{|L{5cm}|C{4cm}|C{4cm}|C{4cm}|}
  \hline
  Symmetry of the superconducting gap $\Delta(\mathbf{k})$ & $G(V)$ at $V\to 0$, $G_n\ll G_{Q}$ & $G(V)$ at $V\to0$, $G_n=G_{Q}$, high-symmetry $\mathbf{r}_0$ & $G(V)$ at $V\to0$, $G_n=G_{Q}$, generic $\mathbf{r}_0$\\ \hline
  Trivial representation & $0$ & $2G_{Q}$ & $2G_{Q}$ \\ \hline
  Point-symmetry broken, TRS preserved, nodeless $\Delta(\mathbf{k})$ & $0$ & $2G_{Q}\left(1-\gamma_{R} \left|eV/\Delta\right|^4 \right)$ & $2G_{Q}\left(1-\gamma_{R} \left|eV/\Delta\right|^4 \right)$ \\ \hline
Nontrivial representation, TRS preserved, nodal $\Delta(\mathbf{k})$ & $\sim G_n \left|eV/\Delta\right|$ & $G_{Q} \gamma_V \left|eV/\Delta\right|$ & $G(0) +G_{Q} \gamma_V \left|eV/\Delta\right|$ \\ \hline
  Point-symmetry broken, TRS preserved, nodal $\Delta(\mathbf{k})$ & $\sim G_n \left|eV/\Delta\right|$ & $G(0) +G_{Q}\gamma_V\left|eV/\Delta\right|$ & $G(0) +G_{Q}\gamma_V\left|eV/\Delta\right|$ \\ \hline
  Nontrivial representation, TRS broken & $0$ &  $0$ & $G(0) -G_{Q} \gamma_{C} \left|eV/\Delta\right|^2$ \\ \hline
  Point-symmetry and TRS broken & $0$ & $G(0) -G_{Q}\gamma_C\left|eV/\Delta\right|^2$ & $G(0) -G_{Q}\gamma_C\left|eV/\Delta\right|^2$\\
  \hline
\end{tabular}
\caption{The differential conductance $G(V)=dI(V)/dV$ at a small bias $V\to0$ and strong tunneling $s_0=0$ for different symmetries of the superconducting gap. Here, $G_Q = e^2/(\pi \hbar)$ is the conductance quantum, $G_n=G_{Q}$ is the differential conductance of the ideal contact between tip and 2D system in the normal state, $\mathbf{r}_0$ is the position of STM tip, and the rest of the constants are discussed in Secs.~S~III.A and S~III.B.
}
\label{tab:crossover-table}
\end{table*}

\subsection{S III.A Conductance at small bias}
\label{sec:app-S3-A}

\subsubsection{S III.A.1 Nodeless real-valued gap}
\label{sec:app-S3-A-1}

Let us start with the case of the nodeless TRS-preserving (real-valued) gap. For small bias $|eV|\ll \Delta_{\rm min}$, the expression for the conductance is given in Eq.~(\ref{app-S2-2-Gtot-sub}) and is determined only by $B_{<}$ and $A_{<}$; see Eqs.~(\ref{app-S1-B-BRe-def}) and (\ref{app-S1-B-ARe-def}) for their definitions. The leading nontrivial order expansion in energy for these coefficients is
\begin{equation}
\label{app-S3-A-min-eps}
A_{<} \approx eV \left\langle \frac{|u_{\mathbf{k}}(\mathbf{r}_0)|^2}{\left|\Delta(\mathbf{k})\right|} \right\rangle_{\!\!0} \quad \mbox{and} \quad
B_{<} \approx 1 +\frac{|eV|^2}{2} \left\langle \frac{|u_{\mathbf{k}}(\mathbf{r}_0)|^2}{\left|\Delta(\mathbf{k})\right|^2} \right\rangle_{\!\!0}.
\end{equation}
Expanding the conductance up to the leading nontrivial order in $eV$, we derive
\begin{equation}
\label{app-S3-G-min-eps}
G(V) \approx 2 G_Q\left(1 - \gamma_{R} \left|\frac{eV}{\Delta}\right|^4 \right),
\end{equation}
where
\begin{equation}
\label{app-S3-G-min-esp-gammaR}
\gamma_{R} = \frac{\Delta^4}{4} \left( \left\langle \frac{|u_{\mathbf{k}}(\mathbf{r}_0)|^2}{\left|\Delta(\mathbf{k})\right|} \right\rangle_{\!\!0}^2 - \left\langle \frac{|u_{\mathbf{k}}(\mathbf{r}_0)|^2}{\left|\Delta(\mathbf{k})\right|^2} \right\rangle_{\!\!0}\, \right)\,;
\end{equation}
following the main text, we introduced here $\Delta$ as a characteristic value of $|\Delta({\bf k})|$ to make $\gamma_R$ unitless.
These expressions are presented in the third line in Tab.~\ref{tab:crossover-table}.
As one can see, the conductance reaches maximum $G(0)=2G_Q$ at $V\to0$ and decreases as $\left|eV/\Delta\right|^4$ with $|V|>0$. In the case of the gap belonging to a trivial representation of the symmetry group, {\sl e.g.}, the $s$-wave gap $\Delta(\mathbf{k})=\Delta$, it is easy to show that $\gamma_{R}=0$ and $G(V)=2G_Q$ for all $|eV|<\Delta$, which is given also in the second line in Tab.~\ref{tab:crossover-table}.

\subsubsection{S III.A.2 Nodal gap}
\label{sec:app-S3-A-2}

Let us proceed to the case of a nodal gap which may or may not respect the crystalline symmetry of the 2D material. The small-bias limit allows us to simplify Eq.~(\ref{app-S3-A-3-G-def}). To perform simplifications, we start with elucidating the leading-term asymptotes of $A_>$, $A_<$, $B_<$, and $B_>$.

The factor $\langle\dots\rangle_\varepsilon$ in the expression for $A_>$, see Eq.~(\ref{app-S1-B-ARe-def}), remains finite in the limit $\varepsilon\to 0$. One can check this by linearizing $\Delta({\bf k})$ around the points where $\Delta({\bf k})=0$ and performing the integration over small intervals of the Fermi line defined by the condition $|\Delta({\bf k})|<|\varepsilon|$. As the result, we find the leading asymptote of $A_>$:
\begin{equation}
\label{abolshe}
A_>=|eV|\sum_{j}\frac{|u_{{\bf k}_j}(\mathbf{r}_0)|^2}{2|\Delta^\prime_j|},
\end{equation}
where $\Delta^\prime=\partial\Delta/\partial\tau$ is the derivative of the gap over a dimensionless vector tangential to the Fermi line, and $j$ stands for the $j$-th zero of the gap.

A similar analysis of $B_>$ shows that
\begin{equation}
\label{bbolshe}
B_>\propto (eV)^2.
\end{equation}
Moreover, the proportionality coefficient in the above equation is not zero only if $\Delta({\bf k})$ violates the lattice symmetry, or if ${\bf r}_0$ is not a high-symmetry point.

The factor $\langle\dots\rangle_\varepsilon$ in the expression for $A_<$, see Eq.~(\ref{app-S1-B-ARe-def}), is logarithmically divergent at $\varepsilon\to 0$, so the leading term in the $A_<$ asymptote scales as
\begin{equation}
\label{amenshe}
A_<\propto eV\ln (\Delta/|eV|).
\end{equation}
Here the value of $\Delta$ is inconsequential within the logarithmic accuracy.

Lastly, $B_{<}=0$ if $\Delta({\bf k})$ and ${\bf r}_0$ do not violate the respective symmetries. Otherwise, the leading expansion term of $B_{<}$ is a constant,
\begin{equation}
\label{bmenshe}
B_<(0)=\left\langle |u_{\mathbf{k}}({\bf r}_0)|^2\frac{\Delta^*(\mathbf{k})}{|\Delta(\mathbf{k})|}\right\rangle_0
\end{equation}
Corrections to Eq.~(\ref{bmenshe}) start with a term $\propto (eV)^2\ln(\Delta/|eV|)$.

Turning to the low-bias asymptotic behavior of the differential conductance, we notice that $A_<$ appears in Eq.~(\ref{app-S3-A-3-G-def}) only in combination $A_<B_>B_<(0)$ or as a higher power ($A_<^2$ or $A_<^4$). As it follows from Eqs.~(\ref{abolshe})--(\ref{amenshe}), all these terms are sub-leading with respect to $A_>$ directly appearing in Eq.~(\ref{app-S3-A-3-G-def}). Aiming at the two leading terms of the differential conductance asymptote, we keep only $A_>$ and $B_<(0)$ in Eq.~(\ref{app-S3-A-3-G-def}) and simplify it to:
\begin{equation}
\label{GVsimple}
G(V)=4G_Q\frac{2|B_<(0)|^2 +\left[1+|B_<(0)|^2\right]A_>}{\left[1+|B_<(0)|^2\right]\left[1+|B_<(0)|^2+4A_>\right]}\,.
\end{equation}
This expression is valid up to the first-order expansion in $A_>$; see Eq.~(\ref{abolshe}) for its explicit form. Performing the expansion, we arrive at the final result
\begin{equation}
\label{app-S3-G-fin}
G(V) \approx G(0) + G_{Q}\gamma_{V} \left|\frac{eV}{\Delta}\right|
\end{equation}
with
\begin{equation}
\label{gammaV}
\quad G(0)=8G_Q\frac{|B_<(0)|^2}{\left[1+|B_<(0)|^2\right]^2},\quad \gamma_V=2\Delta\frac{1-6|B_<(0)|^2+|B_<(0)|^4}{\left[1+|B_<(0)|^2\right]^3}\sum_{j}\frac{|u({\bf k}_j)|^2}{|\Delta^\prime_j|}\,.
\end{equation}
We introduced the gap scale $\Delta$ here to make $\gamma_V$ dimensionless and conform with the notations of the main text and of Tab.~\ref{tab:crossover-table}. The obtained here form of $G(0)$ agrees with Eq.~(15) in the main text and exemplifies the symmetry analysis conclusions therein. The results for real- and complex-valued gaps are summarized in the fourth, fifth, and sixth lines of Tab.~\ref{tab:crossover-table}.

It is notable that the slope of the conductance $\propto\gamma_{V}$ in Eq.~(\ref{app-S3-G-fin}) changes sign at $|B_{\rm crit}(0)|^2=3-2\sqrt{2} \approx 0.17$; see Eq.~(\ref{gammaV}). Therefore, depending on the structure of the gap and the Bloch functions, differential conductance can grow or decay with $|V|$. This change of the slope correlates with the zero-bias conductance: $\gamma_V>0$ at $G(0) \leq G_Q$ for $|B(0)|\leq |B_{\rm crit}(0)|$, and $\gamma_V<0$ at $G_Q \leq G(0) \leq 2G_Q$; see also Fig.~\ref{fig:app-S3-C-Delta0=0.45} for numerical results.
Finally, we notice that the linear scaling with $V$ does not hold for gaps, where both the gap and its derivatives vanish at the nodal points.

\subsubsection{S III.A.3 Nodeless complex-valued gap}
\label{sec:app-S3-A-3}

Finally, let us consider the case of the nodeless TRS-breaking (complex-valued) gap. In the case of subgap energies, $|eV/\Delta|<1$, the conductance is given in Eq.~(\ref{app-S2-2-Gtot-sub}), where the defined in Eqs.~(\ref{app-S1-B-BRe-def}) and (\ref{app-S1-B-ARe-def}) coefficients $A_{<}$ and $B_{<}$ in the leading nontrivial order in $|eV/\Delta|$ are
\begin{equation}
\label{app-S3-A-3-A-min-eps}
A_{<} \approx eV \left\langle\frac{|u_{\mathbf{k}}(\mathbf{r}_0)|^2}{\left|\Delta(\mathbf{k})\right|} \right\rangle_{0} \quad \mbox{and} \quad B_{<} \approx B_<(0)+ \frac{|eV|^2}{2} \left\langle \frac{\Delta^{*}(\mathbf{k})}{|\Delta(\mathbf{k})|} \frac{|u_{\mathbf{k}}(\mathbf{r}_0)|^2}{|\Delta(\mathbf{k})|^2} \right\rangle_{0}
\end{equation}
with $B_<(0)$ given in Eq.~(\ref{bmenshe}). Therefore, the leading nontrivial order result for the conductance is
\begin{equation}
\label{app-S3-A-3-G-min-eps}
G(V) \approx G(0) - G_{Q}\gamma_{C} \left|\frac{eV}{\Delta}\right|^2,
\end{equation}
where
\begin{align}
\label{app-S3-A-3-G0}
\gamma_{C} &=  \frac{1-|B_<(0)|^2}{\left[1+|B_<(0)|^2\right]^4} \left\{\tilde{a}^2 |B_<(0)|^2 -\left[1+|B_<(0)|^2\right]{\rm Re}{\left[B_<(0) \tilde{b}^{*}\right]}\right\},\\
\label{app-S3-A-3-b0-a1}
\quad \tilde{a} &= 4|\Delta| \left\langle\frac{|u_{\mathbf{k}}(\mathbf{r}_0)|^2}{\left|\Delta(\mathbf{k})\right|} \right\rangle_{0}, \quad \quad \quad \quad \tilde{b} = 8\Delta^2 \left\langle\frac{\Delta^{*}(\mathbf{k})}{|\Delta(\mathbf{k})|}\frac{|u_{\mathbf{k}}(\mathbf{r}_0)|^2}{|\Delta(\mathbf{k})|^2} \right\rangle_{0}\,,
\end{align}
and $G(0)$ is defined in Eq.~(\ref{gammaV}). The results (\ref{app-S3-A-3-G-min-eps}) and (\ref{app-S3-A-3-G0}) are given in the seventh line of Tab.~\ref{tab:crossover-table}.

Compared to the conductance for the nodeless real-valued gap in Eq.~(\ref{app-S3-G-min-eps}), we have a different dependence on $|eV|$, {\sl i.e.}, $(eV)^2$ vs. $(eV)^4$. The zero-bias conductance may be less than $2 G_Q$. As discussed in the main text, $G(V)$ is identically zero below a nodeless gap breaking TRS and belonging to a nontrivial representation of the crystalline symmetry group.

\subsection{S III.B Conductance near van Hove singularities}
\label{sec:app-S3-B}

The extrema of $\Delta({\bf k})$ function give rise to van Hove singularities in the density of states (DOS) of Bogoliubov quasiparticles. At weak tunneling, singularities of DOS lead to sharp peaks in the differential conductance at corresponding biases. In the strong-tunneling limit ($s_0=0$) these peaks may transform into singular minima.

This effect is prominent in the case of tunneling into a high-symmetry point $\mathbf{r}_0$ of a superconductor with a nodal real-valued gap respecting the lattice symmetry. In this case, $a_p=a_h=0$ and, therefore, $r_{ph}=r_{hp}=0$. On the contrary, $a_{ph}$ is logarithmically divergent at van Hove singularities, {\sl e.g.},
\begin{equation}
\label{vHsym1}
a_{ph}\propto\ln\left|\frac{\Delta_{\rm max}}{\Delta_{\rm max}-|eV|}\right|\,.
\end{equation}
This singularity determines the behavior of $r_p$, $r_h$, and ultimately $G(V)$ at biases approaching the van Hove singularity points. Using Eq.~(\ref{vHsym1}) in Eq.~(\ref{app-S1-B-r-pp}), we easily find that
\begin{equation}
\label{vHsym2}
1-r_p\propto \left\{\ln\left|\frac{\Delta_{\rm max}}{\Delta_{\rm max}-|eV|}\right|\right\}^{-1}
\end{equation}
at $s_0=0$. Substituting this asymptote into Eq.~(\ref{app-S1-B-Gtot-def}), we find
\begin{equation}
\label{vHsym3}
G(V)\propto \left\{\ln\left|\frac{\Delta_{\rm max}}{\Delta_{\rm max}-|eV|}\right|\right\}^{-1}\,
\end{equation}
{\sl i.e.}, the differential conductance reaches zero at the minimum.

If $\Delta({\bf k})$ breaks the point symmetry or ${\bf r}_0$ deviates from a high-symmetry point, then van Hove divergence occurs in $a_p$ and $a_{ph}$ at the same energies. As the result, the leading $\propto\ln^2(\dots)$ divergence of $a_{ph}^2-a_pa_h$ in the denominators of the expressions for the amplitudes $r_p$ and $r_{ph}$ as well as in the numerator of $r_p$ cancels out, see Eqs.~(\ref{app-S1-B-rph-fin}) and (\ref{app-S1-B-r-pp}) at $s_0=0$. There are two consequences of this cancellation. First, the conductance remains finite at van Hove singularities. Second, $G(V)$ may display a discontinuity at the singular point. The discontinuity comes from the step-function contribution to $a_p$ and $a_{ph}$ accompanying the logarithmically-divergent terms of the type~(\ref{vHsym1}).

\subsection{S III.C Numerical results}
\label{sec:app-S3-C}

In this Section, we supplement the results in the main text and visualize the conductance (\ref{app-S3-A-3-G-def}) for a few types of the gap that were not shown there. We assume a circular Fermi surface parameterized by the angle $\varphi$ and use the combined $s+d$ real gap $\Delta(\varphi)=\Delta_s+\Delta_d \cos{(2\varphi)}$ as a representative example. Depending on the relation between $\Delta_s$ and $\Delta_d$, this gap (i) belongs to a trivial representation at $\Delta_d=0$, (ii) belongs to a nontrivial representation and is nodal at $\Delta_s=0$, (iii) breaks the point symmetry and is nodal at $\Delta_s<\Delta_d$, (iv) breaks the point symmetry and is nodeless at $\Delta_s>\Delta_d$. In addition, we consider an example of a TRS-broken gap. In our numerical calculations, we set $u_{\mathbf{k}}(\mathbf{r}_0)=1$.

\begin{figure*}[!ht]
\centering
\subfigure{\includegraphics[width=0.31\textwidth]{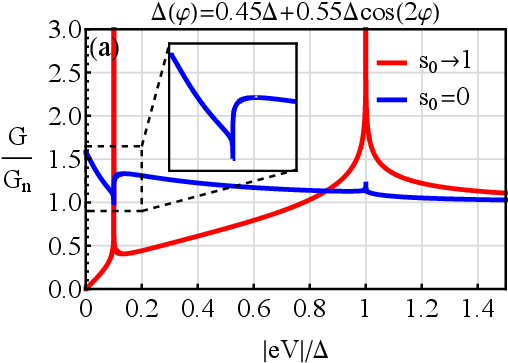}}
\hspace{0.01\textwidth}
\subfigure{\includegraphics[width=0.31\textwidth]{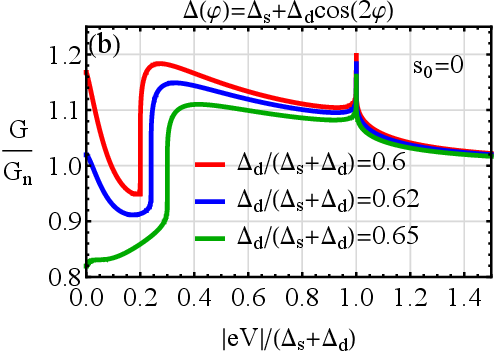}}
\hspace{0.01\textwidth}
\subfigure{\includegraphics[width=0.31\textwidth]{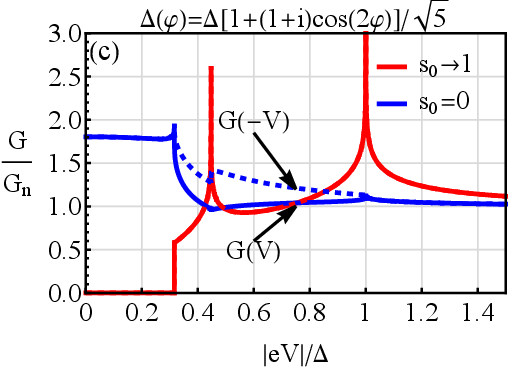}}
\caption{
The dependence of the normalized conductance $G/G_{n}$ on bias $|eV|/\Delta$. We use $\Delta(\varphi) = \Delta_s + \Delta_d \cos{(2\varphi)}$ with $\Delta=\Delta_s+\Delta_d$ and $\Delta_d/\Delta=0.55$ in (a) and $\Delta_d/\Delta=0.6,0.62,0.65$ in (b). The conductance for the TRS-breaking gap  $\Delta(\varphi) = \Delta \left[1 +(1+i)\cos{(2\varphi)}\right]/\sqrt{5}$ is shown in (c). The particle-hole symmetry is broken at $\Delta_{\rm min} < |eV| < \Delta_{\rm max}$, see solid and dashed lines. In all panels, we assume an isotropic electron dispersion relation and use $u_{\mathbf{k}}(\mathbf{r}_0)=1$ in Eqs.~(\ref{app-S1-B-BRe-def}), (\ref{app-S1-B-ARe-def}), and (\ref{app-S3-A-3-G-def}).
}
\label{fig:app-S3-C-Delta0=0.45}
\end{figure*}

For the gap with a broken point symmetry, $\Delta(\varphi)=\Delta_s+\Delta_d\cos{(2\varphi)}$ with $\Delta_s\neq 0$ and $\Delta_d\neq 0$, the nodal gap is realized for $|\Delta_s|<|\Delta_d|$ and is shown in Fig.~\ref{fig:app-S3-C-Delta0=0.45}(a). The behavior of the conductance at $V\to0$ and $s_0=0$ is described by Eq.~(\ref{app-S3-G-fin}). In this case, $G(0)\neq 0$ even if ${\bf r}_0$ is a high-symmetry point; the sign of the slope changes from negative to positive with the tunneling strength. In addition, since there are two types of gap maxima, {\sl i.e.}, at $|\Delta_s+\Delta_d|$ and $|\Delta_s-\Delta_d|$, we observe two non-analytical features in the conductance. While these features are sharp peaks at weak tunneling $s_0\to1$, they might transform into discontinuities for a strong tunneling $s_0=0$; {\sl cf.} red and blue lines in Fig.~\ref{fig:app-S3-C-Delta0=0.45}(a).

Upon increasing the admixture of the $d$-wave component, the conductance changes its zero-bias value and slope at $V\to0$. As one can see from Fig.~\ref{fig:app-S3-C-Delta0=0.45}(b), it decays with $|eV|$ and has $G_Q\leq G(0)< 2G_Q$ for $\Delta_d/\left(\Delta_s+\Delta_d\right) \lesssim 0.62$; the decay changes to growth for $\Delta_d/\left(\Delta_s+\Delta_d\right) \gtrsim 0.62$. Indeed, according to Eqs.~(\ref{app-S3-G-fin}) and (\ref{gammaV}), the sign of the slope of the conductance changes at $|B_{\rm crit}(0)| = \sqrt{2}-1 \approx 0.41$. For $\Delta(\varphi) = \Delta_s +\Delta_d \cos{(2\varphi)}$, this is equivalent to $\Delta_d/\left(\Delta_s+\Delta_d\right) \approx 0.62$.

Finally, as we discussed in Sec.~II.B, it is possible to have an asymmetric conductance $G(V)\neq G(-V)$ for certain TRS-breaking gaps even at $s_0=0$. We present the conductance for such a gap, $\Delta(\varphi) = \Delta \left[1 +(1+i)\cos{(2\varphi)}\right]/\sqrt{5}$, in Fig.~\ref{fig:app-S3-C-Delta0=0.45}(c). Due to the structure of the term responsible for the particle-hole symmetry breakdown, {\sl i.e.}, $\sign{V}\,\mbox{Im}{\left\{B_{<}\,B_{>}^{*}\right\}}$ in Eq.~(\ref{app-S3-A-3-G-def}), the asymmetry appears only for intermediate values of the bias such that $\Delta_{\rm min} < |eV| < \Delta_{\rm max}$; see the solid and dashed lines in Fig.~\ref{fig:app-S3-C-Delta0=0.45}(c).

\bibliography{library-short}